\documentclass[prb]{revtex4}
\newcommand{\lb}{\label}

\begin{document}
\title{Superconductivity in the Model with non Cooper Pairs}
\author{I.M.Yurin, V.B.Kalinin}
\affiliation{Institute of Physical Chemistry, Leninskiy prosp. 31,
GSP-1, Moscow, 119991, Russia} \email{yurinoffice@aport2000.ru}

\begin{abstract}
On a basis of earlier substantiated expression for effective
potential of electron-electron attraction in metals the assumption
of an opportunity of formation of classically bound pairs is put
forward. It was shown that in distinction from the Cooper pair,
the energy of the electron pair in this approach is negative in
the centrum of mass of the pair. This fact changes the pattern of
the superconductivity phenomenon in the proposed model. First, the
gap in the one--particle spectrum appears due to the effect of a
"mean field" on the energy of electron state from the side of
occupied states, the nearest neighbors over the momenta grid.
Second, the condensate is formed by electron states with energies
below that of the gap edge. No Bose condensation does occur in the
system proceeding from the proposed model. In quasi-classical
approximation Londons' hypothesis is substantiated.
\end{abstract}


\maketitle

\section{Introduction}
At present, it is the superconductivity theory put forward by
Bardeen, Cooper and Schrieffer in 1957 that is considered
generally accepted ~\cite{bib-1}. In the framework of this theory,
one succeeded in clarifying a great many of experiments devoted to
investigating so-called conventional superconductors, i.e., those with
rather low transition temperatures. At the time of discovering
high-temperature superconductivity (HTSC), the most optimistic
assessments of BCS theory yielded the value of about 40 K as the
possible maximal transition temperature $T_S$.

The discovery of
HTSC divided research workers into two big groups. One part of
research workers is continuing to consider the BCS theory to be
suitable for describing HTSC: for this purpose, it would be
sufficient to substantiate the introduction of large values of
constants of electron-phonon interaction. Now, other research
workers consider the BCS theory to be applicable to conventional superconductivity.
However, for example, in cuprates one has to search for novel
mechanisms of the pairing of electrons. As concerns the authors of
the present paper, they are in obvious minority, forming a group
considering it to be necessary to radically revise the existing
superconductivity theory. In the framework of the novel theory
that is being developed in our works the HTSC has been naturally
substantiated, whereas the descriptions of superconductivity and
HTSC are lacking essential differences.

In order to properly understand the present-day state of the
superconductivity theory, it is needed to thoroughly consider the
generally accepted BCS theory. The BCS Hamiltonian binds the pairs
with zero momenta only, and has the following form:
\begin{eqnarray}
 H_{BCS}  = \sum\limits_{\sigma ,{\bf k}} {T_{\bf
k} c_{\sigma ,{\bf k}}^ +  c_{\sigma ,{\bf k}} }  - \frac{g}
{\Omega }\sum\limits_{\scriptstyle E_F  - \omega _D  < T_{\bf k}
< E_F  + \omega _D  \hfill \atop
  \scriptstyle E_F  - \omega _D  < T_{\bf p}  < E_F  + \omega _D  \hfill}  {c_{ \uparrow ,{\bf p}}^ +  c_{ \downarrow , - {\bf p}}^ +  c_{ \downarrow , - {\bf k}} c_{ \uparrow ,{\bf k}} } ,
\lb{1}
\end{eqnarray}
where $T_{\bf k}$ is the kinetic energy, $c_{\sigma ,{\bf k}}^+$
and $c_{\sigma ,{\bf k}}$ are the operators of the creation and
annihilation of the electron with momentum ${\bf k}$ and spin
index $\sigma = \pm \frac{1} {2}$(or $\sigma  =  \uparrow ,
\downarrow$ respectively), $E_F$ is the Fermi energy, $\Omega$ is
the volume of the system, and $\omega _D$ is the Debye frequency.

A rigorous analysis shows that Hamiltonian ~(\ref{1}) describes
a dielectric rather than a superconductor. It is possible to
become convinced in that already by scrutinizing so-called
"elementary Cooper superconductor", which is essentially a single
pair of electrons bound together by the interaction term of the
Hamiltonian (\ref{1}) (it is curious that in distinction from
the classically bound pairs, the energy of Cooper pair in the
system of the centrum of masses with realistically chosen
parameters of the model is positive).

Let us assume for the sake of simplicity that $\omega _D  = E_F$.
If we apply to the Cooper superconductor a voltage $V$ of such a
value that the condition $2eV < E_b$ is fulfilled, where $E_b$ is
the energy of a pair to be counted from the bottom of the
conductivity zone and $e$ is the electron charge,  then the
response of such a system to the electrical field will be
calculated in the framework of the perturbation theory, for the
energy of the Cooper pair is separated from all other states by an
energy gap $E_b$. Thus, the Cooper "superconductor" is actually
incapable of screening the electrical field, that is, in the BCS
system even the properties of normal metal have been lost.

It may seem that Bose condensation for many pairs can
substantially change the situation. This supposition refutes the
strict solution of a many electron problem for the reduced BCS
Hamiltonian, as given in study ~\cite{bib-2}. It turns out that the
ground state of such a system is also separated from other states
by an energy gap. Therefore, in order to induce current, it is
necessary to consume a finite quantity of energy, which is
incompatible with the claimed superconductivity properties. In
other words, the BCS model actually describes the Mott
hypothetical transition, when the ground state of the system
proves to be dielectric, and has nothing in common with the
superconductivity problem.

It may seem that we have to deal with a shortcoming of the
original formulation of BCS theory. In the further course, when we
pass over to considering the bound pairs with the momenta that are
different from zero, this shortcoming disappears. Let us show this
does not correspond to reality.

So, in the framework of Gor'kov transformation ~\cite{bib-3}, some
of the interaction terms of the Hamiltonian
\begin{eqnarray}
H = \sum\limits_{\sigma ,{\bf k}} {T_{\bf k} c_{\sigma ,{\bf k}}^
+  c_{\sigma ,{\bf k}} }  + \sum\limits_{\sigma ,\nu }
{\sum\limits_{{\bf p},{\bf k},{\bf q}} {V_{\bf q}^{real} c_{\sigma
,{\bf p} + {\bf q}}^ +  c_{\nu ,{\bf k} - {\bf q}}^ +  c_{\nu
,{\bf k}} c_{\sigma ,{\bf p}} } }
\lb{2}
\end{eqnarray}
are diagonalized by a transition to the operators of the creation
and annihilation of bogolons:
\begin{eqnarray}
  &\gamma _{\sigma ,{\bf p}}^ +   = u_p c_{\sigma ,{\bf p}}^ +   - 2\sigma v_p c_{ - \sigma , - {\bf p}}, \cr
  &\gamma _{\sigma ,{\bf p}}  = u_p c_{\sigma ,{\bf p}}  - 2\sigma v_p c_{ - \sigma , - {\bf p}}^ +  .\hfill
\lb{3}
\end{eqnarray}

The parameters $u_p$ and $v_p$ are determined in the framework of
the self-consistent Hartree--Fock procedure based on equation for
the bogolons' spectrum $ \left[ {H,\gamma _{\sigma ,{\bf p}}^ + }
\right] = E_p \gamma _{\sigma ,{\bf p}}^ + $. The appearance of
the gap in the spectrum is treated as a transition into the
superconductivity state.

Now we are going to note that equation $ \left[ {H,\gamma _{\sigma
,{\bf p}}^ +  } \right] = E_p \gamma _{\sigma ,{\bf p}}^ + $ has a
physical sense only in the case where the vacuum state $\left| {BCS}
\right\rangle$, satisfying the condition $\gamma _{\sigma ,{\bf
p}} \left| {BCS} \right\rangle  = 0$, is stationary (in the BCS
theory this vacuum state is considered the ground state of the system). For
further purpose, it is sufficient to write the Hamiltonian
(\ref{2}) in the operators $\gamma _{\sigma ,\mathbf{p}}^+$ and
$\gamma _{\sigma ,\mathbf{p}}$.

In the case of the BCS Hamiltonian accounting for the terms with
the structure $\gamma ^ + \gamma ^ + \gamma ^ + \gamma ^ +$ we
derive the following expression $\left| {\delta \psi } \right|
\sim N/\Omega$ , where  $\delta \psi$ is a correction of the first
order of the perturbation theory to the wave function of the state
$\left| {BCS} \right\rangle$, and $N$ is the mean number of
electrons in the system. But, in a general case for Hamiltonian
(\ref{2}), the analogous terms give $\left| {\delta \psi }
\right| \sim N^{3/2} /\Omega$, or $\left| {\delta \psi } \right|
\sim n\sqrt N$, where $n$ is concentration of electrons in the
system. Therefore, the Gor'kov procedure is definitely incorrect
when describing real systems with large dimensions.

Now, let us consider Eliashberg's equations ~\cite{bib-4}, which
are frequently referred to as the modern formulation of BCS theory
~\cite{bib-5}. These equations were derived from the Fr\"ohlich model
in 1960. At the same time, it is long known ~\cite{bib-6} that in
the Fr\"ohlich model it is impracticable to effect the consistent
taking into account of the Coulomb electron-electron interaction
in the procedure of screening electron--phonon potential. This
circumstance reveals the irrelevance of the Eliashberg's equations
for Green's functions of electrons and phonons with respect to the
essential point of the renormalization problem. Moreover, the use
of anomalous expectation values makes the determination of the
ground state of the system as unsubstantiated as in the case of
Gor'kov transformation.

The consistent renormalization in the system of electrons and
phonons was for the first time carried into effect in study
~\cite{bib-7}. It has been shown that introduction of an
additional kinetic terms $\sim q^{ - 1} \left( 2{b_\mathbf{q}^ +
b_\mathbf{q}  - b_\mathbf{q}^ +  b_{ - \mathbf{q}}^ +   -
b_\mathbf{q} b_{ - \mathbf{q}} } \right)$, where $b_\mathbf{q}^+$
and $b_\mathbf{q}$ are the operators of the creation and
annihilation of the phonons of longitudinal polarization with
momentum $\mathbf{q}$, enables one to balance a series of
singularities in the long-wave limit, that arise in the process of
renormalization.

The consideration of the long-wave limit of the electron--phonon
and, accordingly, electron--electron interaction is especially
relevant in studying the electron pairs that are bound in the
classical sense. In fact, it is easy to take into account the
short-wave interaction in the framework of the perturbation theory
because of high values of the arising energy denominators.
Therefore, in distinction from the BCS theory, it is just the
long-wave interaction that should play the crucial role in the
substantiation of physically relevant superconductivity theories.

In the framework of the same works ~\cite{bib-7}, when taking into
consideration electronic correlations, we have derived an
expression for the effective Hamiltonian of degenerated Fermi gas
in the model of monatomic metal:
\begin{eqnarray}
H_{eff}  = \sum\limits_{\sigma ,\mathbf{k}} {T_\mathbf{k}
c_{\sigma ,\mathbf{k}}^ +  c_{\sigma ,\mathbf{k}} }  +
\sum\limits_{\sigma ,\nu }
{\sum\limits_{\mathbf{p},\mathbf{k},\mathbf{q}} {\tilde
U_\mathbf{q}^{\mathbf{p},\mathbf{k}} c_{\sigma ,\mathbf{p} +
\mathbf{q}}^ +  c_{\nu ,\mathbf{k} - \mathbf{q}}^ +  c_{\nu
,\mathbf{k}} c_{\sigma ,\mathbf{p}} } }.
\lb{4}
\end{eqnarray}
In that, with $p,k \gg mS$ , we have
\begin{eqnarray}
\tilde U_\mathbf{q}^{\mathbf{p},\mathbf{k}}  \sim  - \left(
{\frac{{zm}} {{3M}}} \right)^2 \frac{{e^2 }} {{\varepsilon q^2
}}\frac{{K_F^2 }} {{q^2  - 4m^2 s^2 }}\frac{{K_F^2 }} {{q^2  -
4m^2 s_1^2 }},
\lb{5}
\end{eqnarray}
where $z$ is the number of conductivity electrons per cell, $m$ is
the zone mass of electron, $M$ is the mass of ion, $\varepsilon$
is the dielectric permittivity of the electrons of valence zone,
$K_F$ is the Fermi wave vector, $s$ is the initial velocity of
phonons, which is linked with the observed sound velocity  $S$ in
metal by a relationship $S = \sqrt {s^2  + \frac{{zm}} {{3M}}V_F^2
}$, $V_F$ is the Fermi velocity, $s_1^2 = \left( {1 + \frac{{zm}}
{{3M}}\frac{{\varepsilon K_F^2 }} {{\lambda ^2 }}} \right)s^2$,
and $\lambda ^{ - 1}$ is the Thomas-Fermi screening length. In
study ~\cite{bib-7}, we actually considered a doped semiconductor.
In such a case, the electron states may be divided into the states
of valence and conductivity zones. In this, the values of
dielectric permittivity $\varepsilon$ and the initial velocity of
phonons $s$ are of explicit parent dielectric origin. Future
investigations will certainly show how correct is the
approximation suggested above for describing the metals of
periodic system.

Now, we are going to make several remarks pertaining to the
subject-matter of this paper. Firstly, the presence of
singularities in Eq. (\ref{5}) is not at all unavoidable in the
framework of the approach being developed. As a matter of fact,
taking into account the phonon and electron scattering will be
conducive to the disappearance of singularities in Eq. (\ref{5}):
this may be achieved by introducing supplementary terms to the
initial Hamiltonian introduced in ~\cite{bib-7}.

Secondly, one has to perceive distinctly the peculiarities of the
construction of the ground state in the framework of the proposed
approach. From the very beginning, we believe the ground state of
the system may be represented in the following form:
\begin{eqnarray}
\left| {\Phi _0 } \right\rangle  = \prod\limits_{\sigma ,k < K_F }
{C_{\sigma ,\mathbf{k}}^ +  } \left| 0 \right\rangle ,
\lb{6}
\end{eqnarray}
where $C_{\sigma ,\mathbf{k}}^+$ are the operators of the creation
of quasi-particles originating from electrons.

During the renormalization procedure, taking into account electron
correlations, and solving other problems, these operators are
refined; however, even the appearance of the bound states of
Hamiltonian $H_{eff}$ does not change equation (\ref{6}).
This circumstance qualitatively distinguishes our approach from
the most studies, in which the Hamiltonians presupposing the
presence of bound states are considered to be initial ones, and,
accordingly, the determining of the ground state of the system
represents a separate problem.

\section{Construction of the Model Hamiltonian}

The effective electron--electron interaction (\ref{4}) was
obtained in the first order of the perturbation theory with
respect to the small parameter $\tilde U/T$. Let us try to
generalize the obtained results for the case where the
perturbation theory does not work. So, after the completion of the
renormalization procedure, let us consider the following
Hamiltonian:
\begin{eqnarray}
H  = \sum\limits_{\sigma ,\mathbf{k}} {T_\mathbf{k} c_{\sigma
,\mathbf{k}}^ +  c_{\sigma ,\mathbf{k}} }  + \sum\limits_{\sigma
,\nu } {\sum\limits_{\mathbf{p},\mathbf{k},\mathbf{q}} {
U_\mathbf{q}^{\mathbf{p},\mathbf{k}} c_{\sigma ,\mathbf{p} +
\mathbf{q}}^ +  c_{\nu ,\mathbf{k} - \mathbf{q}}^ +  c_{\nu
,\mathbf{k}} c_{\sigma ,\mathbf{p}} } }.
\lb{7}
\end{eqnarray}

The scheme of introducing the concept of effective Hamiltonian is
similar to that set forth in papers ~\cite{bib-7}. At first, we
consider the Hamiltonian $H_{eff}$ that, as one would think, is even
formally not linked with Eq. (7):
\begin{eqnarray}
H_{eff} &=& \sum\limits_{\sigma ,\mathbf{k}} {T_\mathbf{k}
c_{\sigma ,\mathbf{k}}^ +  c_{\sigma ,\mathbf{k}} }
\nonumber\\ &&
+\frac{1}{2}\sum\limits_{\mathbf{p},\mathbf{k},\mathbf{q}} {\tilde
U_{0,\mathbf{q}}^{\mathbf{p},\mathbf{k}} \left( {c_{ \uparrow
,\mathbf{p} + \mathbf{q}}^ +  c_{ \downarrow ,\mathbf{k} -
\mathbf{q}}^ +   - c_{ \downarrow ,\mathbf{p} + \mathbf{q}}^ + c_{
\uparrow ,\mathbf{k} - \mathbf{q}}^ +  } \right)\left( {c_{
\downarrow ,\mathbf{k}} c_{ \uparrow ,\mathbf{p}}  - c_{ \uparrow
,\mathbf{k}} c_{ \downarrow ,\mathbf{p}} } \right)} \nonumber\\ &&
 +\sum\limits_\sigma
{\sum\limits_{\mathbf{p},\mathbf{k},\mathbf{q}} {\tilde
U_{1,\mathbf{q}}^{\mathbf{p},\mathbf{k}} c_{\sigma ,\mathbf{p} +
\mathbf{q}}^ +  c_{\sigma ,\mathbf{k} - \mathbf{q}}^ +  c_{\sigma
,\mathbf{k}} c_{\sigma ,\mathbf{p}} } }
 \nonumber\\ &&
+\frac{1}{2}\sum\limits_{\mathbf{p},\mathbf{k},\mathbf{q}} {\tilde
U_{1,\mathbf{q}}^{\mathbf{p},\mathbf{k}} \left( {c_{ \uparrow
,\mathbf{p} + \mathbf{q}}^ +  c_{ \downarrow ,\mathbf{k} -
\mathbf{q}}^ +   + c_{ \downarrow ,\mathbf{p} + \mathbf{q}}^ + c_{
\uparrow ,\mathbf{k} - \mathbf{q}}^ +  } \right)\left( {c_{
\downarrow ,\mathbf{k}} c_{ \uparrow ,\mathbf{p}}  + c_{ \uparrow
,\mathbf{k}} c_{ \downarrow ,\mathbf{p}} } \right).}
\lb{8}
\end{eqnarray}
Using transformation
\begin{eqnarray}
  C_{\sigma ,\mathbf{p}}^ +   = c_{\sigma ,\mathbf{p}}^ +   + \sum\limits_\nu  {\sum\limits_{\mathbf{k},\mathbf{q}} {\theta _{\sigma ,\nu ,\mathbf{q}}^{\mathbf{p},\mathbf{k}} c_{\sigma ,\mathbf{p} + \mathbf{q}}^ +  c_{\nu ,\mathbf{k} - \mathbf{q}}^ +  c_{\nu ,\mathbf{k}} } }
  ,\cr
  C_{\sigma ,\mathbf{p}}  = c_{\sigma ,\mathbf{p}}  + \sum\limits_\nu  {\sum\limits_{\mathbf{k},\mathbf{q}} {\theta _{\sigma ,\nu ,\mathbf{q}}^{*\mathbf{p},\mathbf{k}} c_{\nu ,\mathbf{k}}^ +  c_{\nu ,\mathbf{k} - \mathbf{q}} c_{\sigma ,\mathbf{p} + \mathbf{q}} } }
  ,\hfill
\lb{9}
\end{eqnarray}
the Hamiltonian $H_{eff}$ is transformed to the following form
\begin{eqnarray}
  H_{eff} &=& \sum\limits_{\sigma ,\mathbf{k}} {T_\mathbf{k} C_{\sigma ,\mathbf{k}}^ +  C_{\sigma ,\mathbf{k}} }
\nonumber\\
&&+ \frac{1}{2}\sum\limits_{\sigma ,\mathbf{p} \ne \mathbf{k}}
{E_1^{\mathbf{p},\mathbf{k}} C_{\sigma ,\mathbf{p}}^ + C_{\sigma
,\mathbf{k}}^ +  C_{\sigma ,\mathbf{k}} C_{\sigma ,\mathbf{p}} }
\nonumber\\
&&+ \frac{1}{4}\sum\limits_{\mathbf{p} \ne \mathbf{k}}
{E_1^{\mathbf{p},\mathbf{k}} \left( {C_{ \uparrow ,\mathbf{p}}^ +
C_{ \downarrow ,\mathbf{k}}^ +   + C_{ \downarrow ,\mathbf{p}}^ +
C_{ \uparrow ,\mathbf{k}}^ +  } \right)\left( {C_{ \downarrow
,\mathbf{k}} C_{ \uparrow ,\mathbf{p}}  + C_{ \uparrow
,\mathbf{k}} C_{ \downarrow ,\mathbf{p}} } \right)}
\nonumber\\
&&+ \sum\limits_\mathbf{p} {E_0^{\mathbf{p},\mathbf{p}} C_{
\uparrow ,\mathbf{p}}^ +  C_{ \downarrow ,\mathbf{p}}^ +  C_{
\downarrow ,\mathbf{p}} C_{ \uparrow ,\mathbf{p}} }
\nonumber\\
&&+ \frac{1}{4}\sum\limits_{\mathbf{p} \ne \mathbf{k}}
{E_0^{\mathbf{p},\mathbf{k}} \left( {C_{ \uparrow ,\mathbf{p}}^ +
C_{ \downarrow ,\mathbf{k}}^ +   - C_{ \downarrow ,\mathbf{p}}^ +
C_{ \uparrow ,\mathbf{k}}^ +  } \right)\left( {C_{ \downarrow
,\mathbf{k}} C_{ \uparrow ,\mathbf{p}}  - C_{ \uparrow
,\mathbf{k}} C_{ \downarrow ,\mathbf{p}} } \right)} .
\lb{10}
\end{eqnarray}

In distinction from study ~\cite{bib-7}, in Eq. (\ref{8}) it has been taken into
account that in higher than the first orders of the perturbation
theory with respect to parameter   the effective potentials for
singlet and triplet pairs may differ from each other. Moreover,
the parameters of transformation (\ref{9}) will be determined, not from
the perturbation theory, but will be reassigned by the expressions
of a more general form:
\begin{eqnarray}
&&\theta _{\sigma ,\sigma ,\mathbf{q}}^{\mathbf{p},\mathbf{k}}  =
\frac{1}{2}\left( {\delta _{\mathbf{k} - \mathbf{p}}^\mathbf{q}  -
\delta _\mathbf{q}^0 } \right) + \chi
_{1,\mathbf{q}}^{\mathbf{p},\mathbf{k}}, \cr &&\theta _{\sigma , -
\sigma ,\mathbf{q}}^{\mathbf{p},\mathbf{k}}  =-\delta
_\mathbf{q}^0  + \chi _{0,\mathbf{q}}^{\mathbf{p},\mathbf{k}}  +
\chi _{1,\mathbf{q}}^{\mathbf{p},\mathbf{k}} ,
\lb{11}
\end{eqnarray}
where  $\delta _{...}^{...}$ is a Kronecker's 3D-symbol, $\chi
_{0,\mathbf{q}}^{\mathbf{p},\mathbf{k}}$ and $\chi
_{1,\mathbf{q}}^{\mathbf{p},\mathbf{k}}$ are defined by equations
on the two-particle eigen states of the Hamiltonian $H_{eff}$ (see Eq. (\ref{8}) and
Appendix):
\begin{eqnarray}
  &&H_{eff} \left| {S_1^{\mathbf{p},\mathbf{k}} } \right\rangle  = \left( {T_\mathbf{p}  + T_\mathbf{k}  + E_1^{\mathbf{p},\mathbf{k}} } \right)\left| {S_1^{\mathbf{p},\mathbf{k}} } \right\rangle
,\cr
&&H_{eff} \left| {S_0^{\mathbf{p},\mathbf{k}} } \right\rangle
= \left( {T_\mathbf{p}  + T_\mathbf{k}  +
E_0^{\mathbf{p},\mathbf{k}} } \right)\left|
{S_0^{\mathbf{p},\mathbf{k}} } \right\rangle ,
\lb{12}
\end{eqnarray}
where  $\left| {S_1^{\mathbf{p},\mathbf{k}} } \right\rangle  =
\sum\limits_\mathbf{q} {\chi
_{1,\mathbf{q}}^{\mathbf{p},\mathbf{k}} c_{ \uparrow ,\mathbf{p} +
\mathbf{q}}^ +  c_{ \uparrow ,\mathbf{k} - \mathbf{q}}^ +  }
\left| 0 \right\rangle$
 and $\left| {S_0^{\mathbf{p},\mathbf{k}} } \right\rangle  =
\sum\limits_\mathbf{q} {\chi
_{0,\mathbf{q}}^{\mathbf{p},\mathbf{k}} \left( {c_{ \uparrow
,\mathbf{p} + \mathbf{q}}^ +  c_{ \downarrow ,\mathbf{k} -
\mathbf{q}}^ +   - c_{ \downarrow ,\mathbf{p} + \mathbf{q}}^ + c_{
\uparrow ,\mathbf{k} - \mathbf{q}}^ +  } \right)} \left| 0
\right\rangle $.

The equations on the effective potentials  $\tilde
U_{i,\mathbf{q}}^{\mathbf{p},\mathbf{k}}$ are derived in the
following manner. Formally, the transformation is introduced,
which is reciprocal to transformation (\ref{9}):
\begin{eqnarray}
&& c_{\sigma ,\mathbf{p}}^ +   = C_{\sigma ,\mathbf{p}}^ +   +
\sum\limits_\nu  {\sum\limits_{\mathbf{k},\mathbf{q}} {\theta
_{\sigma ,\nu ,\mathbf{q}}^{*\mathbf{k} - \mathbf{q},\mathbf{p} +
\mathbf{q}} C_{\sigma ,\mathbf{p} + \mathbf{q}}^ +  C_{\nu
,\mathbf{k} - \mathbf{q}}^ +  C_{\nu ,\mathbf{k}} } } , \cr &&
c_{\sigma ,\mathbf{p}}  = C_{\sigma ,\mathbf{p}}  +
\sum\limits_\nu  {\sum\limits_{\mathbf{k},\mathbf{q}} {\theta
_{\sigma ,\nu ,\mathbf{q}}^{\mathbf{k} - \mathbf{q},\mathbf{p} +
\mathbf{q}} C_{\nu ,\mathbf{k}}^ +  C_{\nu ,\mathbf{k} -
\mathbf{q}} C_{\sigma ,\mathbf{p} + \mathbf{q}} } } .
\lb{13}
\end{eqnarray}
Using Eq. (\ref{13}), the Hamiltonian (\ref{7}) is written in
operators $C^+$ and $C$. In such a written form, the terms appear
in the Hamiltonian having the structure $C^+C^+C^+CCC$, which, in
the random phase approximation can be reduced to the form
$\left\langle {C^ +  C} \right\rangle C^+C^+CC$ (here and further
the broken brackets  $\left\langle {...} \right\rangle$ designate
the averaging with respect to the ground state). These terms and
some other  arising terms with the structure $C^+C^+CC$ define
supplementary potentials $\tilde
D_{0,\mathbf{q}}^{\mathbf{p},\mathbf{k}}$ and $\tilde
D_{1,\mathbf{q}}^{\mathbf{p},\mathbf{k}}$. Then the
self-consistency equations on the effective potentials  $\tilde
U_{i,\mathbf{q}}^{\mathbf{p},\mathbf{k}}$ become
\begin{eqnarray}
\tilde U_{i,\mathbf{q}}^{\mathbf{p},\mathbf{k}}  =
U_\mathbf{q}^{\mathbf{p},\mathbf{k}}  + \tilde
D_{i,\mathbf{q}}^{\mathbf{p},\mathbf{k}} \lb{14}
\end{eqnarray}
In the case where the perturbation theory is applicable, i.e.,
\begin{eqnarray}
&&\chi _{0,\mathbf{q}}^{\mathbf{p},\mathbf{k}}  \approx \frac{1}
{2}\left( {\delta _\mathbf{q}^0  + \delta _{\mathbf{k} -
\mathbf{p}}^\mathbf{q} } \right),\cr
&&\chi
_{1,\mathbf{q}}^{\mathbf{p},\mathbf{k}}  \approx \frac{1}
{2}\left( {\delta _\mathbf{q}^0  - \delta _{\mathbf{k} -
\mathbf{p}}^\mathbf{q} } \right),
\lb{15}
\end{eqnarray}
Eqs. (\ref{14}) little differ from those derived in study \cite{bib-7}. We are
now interested in the case, where some solutions form the states
of discrete spectrum, and relationships (\ref{15}) corresponding to
these states are not fulfilled.

Let us restrict ourselves to the consideration of the case where
the Hamiltonian has one bound-state solution with a zero spin for
pairs with momentum $\mathbf{P}$. For all the remaining unbound
pairs with $\mathbf{p} + \mathbf{k} = \mathbf{P}$, the following
equation is fulfilled: $\chi
_{0,\mathbf{q}}^{\mathbf{p},\mathbf{k}}  \approx \frac{1}
{2}\left( {\delta _\mathbf{q}^0  + \delta _{\mathbf{k} -
\mathbf{p}}^\mathbf{q} } \right)$. Determine for which pair this
relationship has not been fulfilled (it is the pair forming the
discrete spectrum).

For this purpose, note once more that it is the long-wave part of
the electron--electron interaction, which plays the main role in
the formation of bound states. When considering the
electron--electron interaction of two-particle states in the
framework of the perturbation theory, the energy denominator $\sim
(\mathbf{p} - \mathbf{k})\mathbf{q}$. Hence, it attains the
minimal value when $\mathbf{p} \approx \mathbf{k}$. That is why
the bound pair originates from the state of two free electrons
that are located on one site of the momentum grid or on the
neighboring ones.

We will reckon that the pair binding energy $E_b$ considerably
exceeds corrections $E_i^{\mathbf{p},\mathbf{k}}$  for the pairs
whose spectrum is well described by the perturbation theory. Then,
for a crystal with dimensions $L \times L \times L$  and periodic
boundary conditions, we arrive to the following model Hamiltonian:
\begin{eqnarray}
   H_{mod}  &=& \sum\limits_{\sigma ,\mathbf{k}} {T_\mathbf{k} C_{\sigma ,\mathbf{k}}^ +  C_{\sigma ,\mathbf{k}} }\
\nonumber \\ &&- \sum\limits_\mathbf{k} {E_b C_{ \uparrow ,\mathbf{k}}^ +  C_{ \downarrow ,\mathbf{k}}^ +  C_{ \downarrow ,\mathbf{k}} C_{ \uparrow ,\mathbf{k}} }
\nonumber \\ &&- \frac{1}{4}\sum {E_b \left( {C_{ \uparrow ,\mathbf{p}}^ +  C_{ \downarrow ,\mathbf{k}}^ +   - C_{ \downarrow ,\mathbf{p}}^ +  C_{ \uparrow ,\mathbf{k}}^ +  } \right)\left( {C_{ \downarrow ,\mathbf{k}} C_{ \uparrow ,\mathbf{p}}  - C_{ \uparrow ,\mathbf{k}} C_{ \downarrow ,\mathbf{p}} } \right)}
\nonumber \\ &&  \left\{ {\mathbf{p} \ne \mathbf{k};\left| {\mathbf{p}_\alpha   - \mathbf{k}_\alpha  } \right| \le 2\pi /L,\alpha  = x,y,z} \right\}.
\lb{16}
\end{eqnarray}
The energy of a quasi-particle
 $\tilde E_\mathbf{p}$ in the mean field approximation accounting for
the nearest neighbors over the momenta grid is determined by the following equation:
\begin{eqnarray}
\tilde E_\mathbf{p} C_{\sigma ,\mathbf{k}}^ +   = \left[ {H_{mod} ,C_{\sigma ,\mathbf{k}}^ +  } \right] = \left( {T_\mathbf{k}  - 14E_b \rho _\mathbf{k} } \right)C_{\sigma ,\mathbf{k}}^ +  ,
\lb{17}
\end{eqnarray}
 where $\rho _\mathbf{k}$ is the mean value of the occupation number in the
 vicinity of the site $ \mathbf{k}$.

Eq. (\ref{17}) enables one to obtain the qualitative picture of changes in the electron
spectrum during variation in temperature ~\cite{bib-8}. At the lowest temperatures,
the spectrum has a gap of the magnitude $2\Delta _0  = 14E_b$. As temperature rises,
the gap decreases. At the transition point, the spectrum reveals a singularity,
whereas at a temperature which is higher than the transition point, the gap
completely disappears.

The transition temperature can be estimated from an equation
$\left. {\partial \rho /\partial E} \right|_{\tilde E = \mu }  =  - \infty$,
where  $\mu$ is the chemical potential of electrons. Assuming for $\rho$
the Fermi--Dirac distribution function, we obtain for the transition temperature  $T_s$:
\begin{eqnarray}
T_s  = \frac{{\Delta _0 }}
{2}.
\lb{18}
\end{eqnarray}

In distinction from the BCS model, the gap forming in the system is evidently
superconducting. It follows from the following speculations: firstly, it is
possible to obtain the current state of the system by shifting the wave function
of the ground state in the momenta space, without consuming the finite energy;
secondly, the collision integral in such a system is frozen out according to the
exponential law.

The superconducting condensate forms the ground state of the system.
The thermally excited electrons outside the momenta space, which is occupied by
condensate, and the holes inside the condensate form normal gas that, contrary
to the BCS theory, is electrically neutral.  All the processes of scattering in
the system occur only for the normal gas. Therefore, it proves to be extremely
difficult to estimate the condensate relaxation time with non-zero current.
However, it is most likely that here we have to do with astronomic values.

Winding up the description of the procedure of introducing the model Hamiltonian,
we note once more that in the case where two--particle solutions of discrete spectrum
appear in the system, the transformations (\ref{9}) and (\ref{13}) are of formal, auxiliary
character for the determination of the model Hamiltonian in the space of wave
functions of eigen states of the Hamiltonian $H_{eff}$. These transformations
allow one to derive also expressions for the operators of the values observed,
corresponding to the Hamiltonian $H_{mod}$. Thus, for the operator
$c_{\sigma ,{\bf p}}^ +  c_{\sigma ,{\bf p} + {\bf l}}$  , to be applied to
determining the current density, we have the following expression:
\begin{eqnarray}
  c_{\sigma ,{\bf p}}^ +  c_{\sigma ,{\bf p} + {\bf l}}  &=& C_{\sigma ,{\bf p}}^ +  C_{\sigma ,{\bf p} + {\bf l}}
  \nonumber \\ &&  - \sum\limits_{\nu ,{\bf k}} {C_{\sigma ,{\bf p}}^ +  C_{\nu ,{\bf k}}^ +  C_{\nu ,{\bf k}} C_{\sigma ,{\bf p} + {\bf l}} }
  \nonumber \\ &&  + \sum\limits_{\nu ,{\bf k}} {\sum\limits_{{\bf z},{\bf y}} {\chi _{1,{\bf z} + {\bf y}}^{*{\bf k},{\bf p}} \chi _{1,{\bf y}}^{{\bf k} + {\bf z},{\bf p} - {\bf z} + {\bf l}} C_{\sigma ,{\bf p}}^ +  C_{\nu ,{\bf k}}^ +  C_{\nu ,{\bf k} + {\bf z}} C_{\sigma ,{\bf p} - {\bf z} + {\bf l}} } }
  \nonumber \\ &&  + \sum\limits_{\nu ,{\bf k}} {\sum\limits_{{\bf z},{\bf y}} {\chi _{0,{\bf z} + {\bf y}}^{*{\bf k},{\bf p}} \chi _{0,{\bf y}}^{{\bf k} + {\bf z},{\bf p} - {\bf z} + {\bf l}} C_{\sigma ,{\bf p}}^ +  C_{ - \sigma ,{\bf k}}^ +  C_{ - \sigma ,{\bf k} + {\bf z}} C_{\sigma ,{\bf p} - {\bf z} + {\bf l}} } } .
\lb{19}
\end{eqnarray}
We shall employ this equation to calculate the response of the condensate to
electromagnetic field.

\section{The Behavior of Condensate in Electromagnetic Field}

The appearance of electromagnetic field in the system under consideration is described
by introducing a supplementary term $H_{em}$. In this, the total Hamiltonian of the
system has the following form:
\begin{eqnarray}
H_{tot}  = H_{mod}  + H_{em} ,
\lb{20}
\end{eqnarray}
where
\begin{eqnarray}
&& H_{em}  =  - \frac{1}{c}\sum\limits_{\bf q} {{\bf A}_{ - {\bf q},t} {\bf j}_{\bf q} } ,  \cr
&& {\bf A}_{{\bf q},t}  = L^{ - 3/2} \int {{\bf A}\left( {{\bf r},t} \right)e^{ - i{\bf qr}} d{\bf r}} ,  \cr
&& {\bf j}_{\bf q}  =  - \frac{{eL^{ - 3/2} }}
{{2m}}\sum\limits_{\sigma ,{\bf k}} {\left( {2{\bf k} + {\bf q}} \right)c_{\sigma ,{\bf k}}^ +  c_{\sigma ,{\bf k} + {\bf q}} }  - \frac{{e^2 L^{ - 3} }}
{{mc}}\sum\limits_{\sigma ,{\bf k},{\bf l}} {{\bf A}_{{\bf q} - {\bf l},t} c_{\sigma ,{\bf k}}^ +  c_{\sigma ,{\bf k} + {\bf l}} } ,
\lb{21}
\end{eqnarray}
while the expressions for $c_{\sigma ,{\bf k}}^ +  c_{\sigma ,{\bf k} + {\bf q}}$
and $c_{\sigma ,{\bf k}}^ +  c_{\sigma ,{\bf k} + {\bf l}}$
may be substituted from Eq. (\ref{19}).

We will search for the wave function of the condensate coinciding the wave function of
the ground state of the system in the following form:
\begin{eqnarray}
\left| {\Phi _t } \right\rangle  = \prod\limits_{\sigma ,k < K_F } {\tilde C_{\sigma ,{\bf k}}^ +  \left( t \right)} \left| 0 \right\rangle ,
\lb{22}
\end{eqnarray}
whereas for $\tilde C_{\sigma ,{\bf k}}^ +  \left( t \right)$ we assume the fulfillment
of the following relationship:
\begin{eqnarray}
i\frac{{\partial \tilde C_{\sigma ,{\bf k}}^ +  \left( t \right)}}
{{\partial t}} = \left[ {H_{tot} ,\tilde C_{\sigma ,{\bf k}}^ +  \left( t \right)} \right].
\lb{23}
\end{eqnarray}
It is presumably obvious that Schr\"odinger's equation
$i\frac{\partial }{{\partial t}}\left| {\Phi _t } \right\rangle  = H_{tot}\left| {\Phi _t } \right\rangle$
will then be fulfilled, and, accordingly, the evolution of the system with time
will be correctly described. It is also easy to see that one-time operators
$\tilde C^ +$ and $\tilde C$ satisfy the conventional commutation relationships for fermions.

In the first order of the perturbation theory with respect to parameters ${\bf A}/T$
and $E_i^{{\bf p},{\bf k}} /T$, it is possible to derive the following expressions:
\begin{eqnarray}
  \tilde C_{\sigma ,{\bf p}}^ +  \left( t \right) &=& \sum\limits_{\bf x} {\phi _{{\bf x},t}^{\bf p} C_{\sigma ,{\bf x}}^ +  }  + \sum\limits_{{\bf x},{\bf y}} {\sum\limits_{{\bf k},{\bf z}} {\left( {\psi _{{\bf x},{\bf y},t}^{1,{\bf p},{\bf k}}  - \phi _{{\bf x},t}^{\bf p} \phi _{{\bf y},t}^{\bf k} } \right)\phi _{{\bf z},t}^{*{\bf k}} C_{\sigma ,{\bf x}}^ +  C_{\sigma ,{\bf y}}^ +  C_{\sigma ,{\bf z}} } }   \cr
 \nonumber \\ &&+ \sum\limits_{{\bf x},{\bf y}} {\sum\limits_{{\bf k},{\bf z}} {\left( {\psi _{{\bf x},{\bf y},t}^{0,{\bf p},{\bf k}}  + \psi _{{\bf x},{\bf y},t}^{1,{\bf p},{\bf k}}  - \phi _{{\bf x},t}^{\bf p} \phi _{{\bf y},t}^{\bf k} } \right)\phi _{{\bf z},t}^{*{\bf k}} C_{\sigma ,{\bf x}}^ +  C_{ - \sigma ,{\bf y}}^ +  C_{ - \sigma ,{\bf z}} } } ,  \cr
\tilde C_{\sigma ,{\bf p}} \left( t \right) &=& \sum\limits_{\bf x} {\phi _{{\bf x},t}^{*{\bf p}} C_{\sigma ,{\bf x}} }  + \sum\limits_{{\bf x},{\bf y}} {\sum\limits_{{\bf k},{\bf z}} {\left( {\psi _{{\bf x},{\bf y},t}^{*1,{\bf p},{\bf k}}  - \phi _{{\bf x},t}^{*{\bf p}} \phi _{{\bf y},t}^{*{\bf k}} } \right)\phi _{{\bf z},t}^{\bf k} C_{\sigma ,{\bf z}}^ +  C_{\sigma ,{\bf y}} C_{\sigma ,{\bf x}} } }   \cr
\nonumber \\ &&+ \sum\limits_{{\bf x},{\bf y}} {\sum\limits_{{\bf k},{\bf z}} {\left( {\psi _{{\bf x},{\bf y},t}^{*0,{\bf p},{\bf k}}  + \psi _{{\bf x},{\bf y},t}^{*1,{\bf p},{\bf k}}  - \phi _{{\bf x},t}^{*{\bf p}} \phi _{{\bf y},t}^{*{\bf k}} } \right)\phi _{{\bf z},t}^{\bf k} C_{ - \sigma ,{\bf z}}^ +  C_{ - \sigma ,{\bf y}} C_{\sigma ,{\bf x}} } } ,
\lb{24}
\end{eqnarray}
whereas for $\phi$ and $\psi$ we assume the fulfillment of the following relationships:
\begin{eqnarray}
  && i\sum\limits_{\bf x} {\frac{{\partial \phi _{{\bf x},t}^{\bf p} }}
{{\partial t}}C_{\sigma ,{\bf x}}^ +  } \left| 0 \right\rangle  = H_{tot} \sum\limits_{\bf x} {\phi _{{\bf x},t}^{\bf p} C_{\sigma ,{\bf x}}^ +  } \left| 0 \right\rangle ,  \cr
  && i\sum\limits_{{\bf x},{\bf y}} {\frac{{\partial \psi _{{\bf x},{\bf y},t}^{0,{\bf p},{\bf k}} }}
{{\partial t}}\left( {C_{\sigma ,{\bf x}}^ +  C_{ - \sigma ,{\bf y}}^ +   - C_{ - \sigma ,{\bf x}}^ +  C_{\sigma ,{\bf y}}^ +  } \right)} \left| 0 \right\rangle  = H_{tot} \sum\limits_{{\bf x},{\bf y}} {\psi _{{\bf x},{\bf y},t}^{0,{\bf p},{\bf k}} \left( {C_{\sigma ,{\bf x}}^ +  C_{ - \sigma ,{\bf y}}^ +   - C_{ - \sigma ,{\bf x}}^ +  C_{\sigma ,{\bf y}}^ +  } \right)} \left| 0 \right\rangle ,  \cr
  && i\sum\limits_{{\bf x},{\bf y}} {\frac{{\partial \psi _{{\bf x},{\bf y},t}^{1,{\bf p},{\bf k}} }}
{{\partial t}}C_{\sigma ,{\bf x}}^ +  C_{\sigma ,{\bf y}}^ +  } \left| 0 \right\rangle  = H_{tot} \sum\limits_{{\bf x},{\bf y}} {\psi _{{\bf x},{\bf y},t}^{1,{\bf p},{\bf k}} C_{\sigma ,{\bf x}}^ +  C_{\sigma ,{\bf y}}^ +  } \left| 0 \right\rangle ,
\lb{25}
\end{eqnarray}

Assuming that it is possible to apply the perturbation theory to solving Eqs. (\ref{25})
with accuracy up to linear terms with respect to ${\bf A}$, we may expect the fulfillment
of relationships:
\begin{eqnarray}
  && \phi _{{\bf x},t}^{\bf p}  = \delta _{\bf x}^{\bf p} e^{ - iT_{\bf p} t} ,  \cr
  && \psi _{{\bf x},{\bf y},t}^{0,{\bf p},{\bf k}}  = \frac{1}
{2}\left( {\delta _{\bf x}^{\bf p} \delta _{\bf y}^{\bf k}  + \delta _{\bf y}^{\bf p} \delta _{\bf x}^{\bf k} } \right)e^{ - i\left( {T_{\bf p}  + T_{\bf k}  + E_0^{{\bf p},{\bf k}} } \right)t} ,  \cr
  && \psi _{{\bf x},{\bf y},t}^{1,{\bf p},{\bf k}}  = \frac{1}
{2}\left( {\delta _{\bf x}^{\bf p} \delta _{\bf y}^{\bf k}  - \delta _{\bf y}^{\bf p} \delta _{\bf x}^{\bf k} } \right)e^{ - i\left( {T_{\bf p}  + T_{\bf k}  + E_1^{{\bf p},{\bf k}} } \right)t} .
\lb{26}
\end{eqnarray}
Now, the idea proper of calculations consists in the following.
If we express the density of current ${\bf j}$ in operators $\tilde C^ +$ and $\tilde C$,
then it will be simple to calculate the mean value $\left\langle {\bf j} \right\rangle$.
For this purpose, we will define the transformation which is reciprocal to that,
given in Eq. (\ref{24}):
\begin{eqnarray}
  C_{\sigma ,{\bf x}}^ +   &=& \sum\limits_{\bf p} {\phi _{{\bf x},t}^{*{\bf p}} \tilde C_{\sigma ,{\bf p}}^ +  }   \cr
  &&  + \sum\limits_{{\bf p},{\bf k}} {\sum\limits_{{\bf y},{\bf z}} {\left( {\psi _{{\bf x},{\bf y},t}^{*1,{\bf p},{\bf k}}  - \phi _{{\bf x},t}^{*{\bf p}} \phi _{{\bf y},t}^{*{\bf k}} } \right)\phi _{{\bf y},t}^{\bf z} \tilde C_{\sigma ,{\bf p}}^ +  \tilde C_{\sigma ,{\bf k}}^ +  \tilde C_{\sigma ,{\bf z}} } }   \cr
  &&  + \sum\limits_{{\bf p},{\bf k}} {\sum\limits_{{\bf y},{\bf z}} {\left( {\psi _{{\bf x},{\bf y},t}^{*0,{\bf p},{\bf k}}  + \psi _{{\bf x},{\bf y},t}^{*1,{\bf p},{\bf k}}  - \phi _{{\bf x},t}^{*{\bf p}} \phi _{{\bf y},t}^{*{\bf k}} } \right)\phi _{{\bf y},t}^{\bf z} \tilde C_{\sigma ,{\bf p}}^ +  \tilde C_{ - \sigma ,{\bf k}}^ +  \tilde C_{ - \sigma ,{\bf z}} } } ,  \cr
  C_{\sigma ,{\bf x}}  &=& \sum\limits_{\bf p} {\phi _{{\bf x},t}^{\bf p} \tilde C_{\sigma ,{\bf p}} }   \cr
  &&  + \sum\limits_{{\bf p},{\bf k}} {\sum\limits_{{\bf y},{\bf z}} {\left( {\psi _{{\bf x},{\bf y},t}^{1,{\bf p},{\bf k}}  - \phi _{{\bf x},t}^{\bf p} \phi _{{\bf y},t}^{\bf k} } \right)\phi _{{\bf y},t}^{*{\bf z}} \tilde C_{\sigma ,{\bf z}}^ +  } } \tilde C_{\sigma ,{\bf k}} \tilde C_{\sigma ,{\bf p}}   \cr
  &&  + \sum\limits_{{\bf p},{\bf k}} {\sum\limits_{{\bf y},{\bf z}} {\left( {\psi _{{\bf x},{\bf y},t}^{0,{\bf p},{\bf k}}  + \psi _{{\bf x},{\bf y},t}^{1,{\bf p},{\bf k}}  - \phi _{{\bf x},t}^{\bf p} \phi _{{\bf y},t}^{\bf k} } \right)\phi _{{\bf y},t}^{*{\bf z}} \tilde C_{ - \sigma ,{\bf z}}^ +  \tilde C_{ - \sigma ,{\bf k}} } } \tilde C_{\sigma ,{\bf p}} ,
\lb{27}
\end{eqnarray}
Then, in the operators $\tilde C^ +$ and $\tilde C$ the expression
$c_{\sigma ,{\bf k}}^ +  c_{\sigma ,{\bf k} + {\bf l}}$, included
in the expression for current density will be as follows:
\begin{eqnarray}
  c_{\sigma ,{\bf p}}^ +  c_{\sigma ,{\bf p} + {\bf l}}  &=& \sum\limits_{{\bf p'},{\bf p''}} {\phi _{{\bf p},t}^{*{\bf p'}} \phi _{{\bf p} + {\bf l},t}^{{\bf p''}} \tilde C_{\sigma ,{\bf p'}}^ +  \tilde C_{\sigma ,{\bf p''}} }   \cr
  &&  - \sum\limits_{\nu ,{\bf k'}} {\sum\limits_{{\bf p'},{\bf p''}} {\phi _{{\bf p},t}^{*{\bf p'}} \phi _{{\bf p} + {\bf l},t}^{{\bf p''}} \tilde C_{\sigma ,{\bf p'}}^ +  \tilde C_{\nu ,{\bf k'}}^ +  \tilde C_{\nu ,{\bf k'}} \tilde C_{\sigma ,{\bf p''}} } }   \cr
  &&  + \sum\limits_{{\bf k},{\bf p},{\bf z}} {\sum\limits_{\bf y} {\chi _{1,{\bf z} + {\bf y}}^{*{\bf k},{\bf p}} \chi _{1,{\bf y}}^{{\bf k} + {\bf z},{\bf p} - {\bf z} + {\bf l}} } \sum\limits_{\nu ,{\bf k'},{\bf k''}} {\sum\limits_{{\bf p'},{\bf p''}} {\psi _{{\bf p},{\bf k},t}^{*1,{\bf p'},{\bf k'}} \psi _{{\bf p} - {\bf z} + {\bf l},{\bf k} + {\bf z},t}^{1,{\bf p''},{\bf k''}} \tilde C_{\sigma ,{\bf p'}}^ +  \tilde C_{\nu ,{\bf k'}}^ +  \tilde C_{\nu ,{\bf k''}} \tilde C_{\sigma ,{\bf p''}} } } }   \cr
  &&  + \sum\limits_{{\bf k},{\bf p},{\bf z}} {\sum\limits_{\bf y} {\chi _{1,{\bf z} + {\bf y}}^{*{\bf k},{\bf p}} \chi _{1,{\bf y}}^{{\bf k} + {\bf z},{\bf p} - {\bf z} + {\bf l}} } } \sum\limits_{\nu ,{\bf k'},{\bf k''}} {\sum\limits_{{\bf p'},{\bf p''}} {\psi _{{\bf p},{\bf k},t}^{*0,{\bf p'},{\bf k'}} \psi _{{\bf p} - {\bf z} + {\bf l},{\bf k} + {\bf z},t}^{0,{\bf p''},{\bf k''}} \tilde C_{\sigma ,{\bf p'}}^ +  \tilde C_{ - \sigma ,{\bf k'}}^ +  \tilde C_{ - \sigma ,{\bf k''}} \tilde C_{\sigma ,{\bf p''}} } }   \cr
  &&  + \sum\limits_{{\bf k},{\bf p},{\bf z}} {\sum\limits_{\bf y} {\chi _{0,{\bf z} + {\bf y}}^{*{\bf k},{\bf p}} \chi _{0,{\bf y}}^{{\bf k} + {\bf z},{\bf p} - {\bf z} + {\bf l}} } } \sum\limits_{\nu ,{\bf k'},{\bf k''}} {\sum\limits_{{\bf p'},{\bf p''}} {\psi _{{\bf p},{\bf k},t}^{*1,{\bf p'},{\bf k'}} \psi _{{\bf p} - {\bf z} + {\bf l},{\bf k} + {\bf z},t}^{1,{\bf p''},{\bf k''}} \tilde C_{\sigma ,{\bf p'}}^ +  \tilde C_{ - \sigma ,{\bf k'}}^ +  \tilde C_{ - \sigma ,{\bf k''}} \tilde C_{\mu ,{\bf p''}} } }   \cr
  &&  + \sum\limits_{{\bf k},{\bf p},{\bf z}} {\sum\limits_{\bf y} {\chi _{0,{\bf z} + {\bf y}}^{*{\bf k},{\bf p}} \chi _{0,{\bf y}}^{{\bf k} + {\bf z},{\bf p} - {\bf z} + {\bf l}} } } \sum\limits_{\nu ,{\bf k'},{\bf k''}} {\sum\limits_{{\bf p'},{\bf p''}} {\psi _{{\bf p},{\bf k},t}^{*0,{\bf p'},{\bf k'}} \psi _{{\bf p} - {\bf z} + {\bf l},{\bf k} + {\bf z},t}^{0,{\bf p''},{\bf k''}} \tilde C_{\sigma ,{\bf p'}}^ +  \tilde C_{ - \sigma ,{\bf k'}}^ +  \tilde C_{ - \sigma ,{\bf k''}} \tilde C_{\sigma ,{\bf p''}} } }
\lb{28}
\end{eqnarray}
Limiting ourselves only to unrelativistic terms in the equation for
$\frac{{\partial {\bf j}_{\bf q} }}{{\partial t}}$
with an accuracy of up to the second order with respect to parameter ${\bf A}$,
it is possible to derive
\begin{eqnarray}
\left\langle {\frac{{\partial {\bf j}_{\bf q} }}
{{\partial t}}} \right\rangle  =  - \frac{{e^2 L^{ - 3} }}
{{mc}}\sum\limits_{\sigma ,{\bf p},{\bf l}} {\frac{{\partial {\bf A}_{{\bf q} - {\bf l}} }}
{{\partial t}}\left\langle {c_{\sigma ,{\bf p}}^ +  c_{\sigma ,{\bf p} + {\bf l}} } \right\rangle }  =  - \frac{{e^2 n}}
{{mc}}\frac{{\partial {\bf A}_{\bf q} }}
{{\partial t}},
\lb{29}
\end{eqnarray}
or
\begin{eqnarray}
\left\langle {{\bf j}_{\bf q} } \right\rangle  =  - \frac{{e^2 n}}
{{mc}}{\bf A}_{\bf q} ,
\lb{30}
\end{eqnarray}
which just constitutes the essential point of Londons' hypothesis.

\section{Conclusions}

In our opinion, it is possible to subdivide the existing, relevant from
the viewpoint of physics studies on the superconductivity theory into two
categories enumerated below.

1.  Systems with Bose--Einstein Condensation

In such works, the Hamiltonians with electron-electron attraction are introduced,
which are considered as initial ones. In this, the boson operators corresponding
to the bound states of electron pairs are used to determine the ground state of
the system.

The studies devoted to dealing with this problem are, as a whole, not systematized.
Therefore, it is very difficult to single out the most significant of them.
It should also be noted that some of the papers anticipate the BCS theory ~\cite{bib-9}.
Therefore, the use of the term "Cooper pairs" seems to be not quite suitable,
especially from the viewpoint of the remarks, given in the introduction to our paper.

With some provision, the modifications of the popular $t - j$ model may be related
to those studies. With a certain selection of parameters of the model, for example
$j > t$, a possibility arises of the formation of bound pairs. The condition $j > t$
is linked with a short-wave character of the interaction in the Hubbard form and
Heisenberg's uncertainty principle.  The authors of these works themselves call
the choice of such parameters as "unphysical" ~\cite{bib-10}.

Now, the introduction of Hamiltonian to the studies devoted to the
Bose condensation in Fermi gas is either unsubstantiated or causes
skepticism in the minds of most physicists ~\cite{bib-11}. In this
sense, we may reckon the authors of these works are so far
investigating hypothetical objects.

The concept of Bose condensate in such systems has a distinctly expressed sense.
That is why the Ginzburg--Landau theory using the wave function of Bose condensate
in the definition of the order parameter of the system is quite valid.

2.  Systems with the Fermi Condensation

In this case, the electron-electron attraction arises only with an effective potential
which is calculated in an assumption that the wave function of the ground state coincides
in its form with the function of the ground state of a normal metal
$\left| {\Phi _0 } \right\rangle  = \prod\limits_{\sigma ,k < K_F } {C_{\sigma ,{\bf k}}^ +  } \left| 0 \right\rangle$.
The question to be settled resides merely in selecting the operators
$C_{\sigma ,{\bf k}}^ +$. The effective potential is calculated in the
framework of refining operators $C_{\sigma ,{\bf k}}^ +$ when solving the
problem of renormalization and accounting for electron-electron correlations.
From this viewpoint, the operators $C_{\sigma ,{\bf k}}^ +$ are linked in a
self--consistent manner with the effective potential of the system, and there
is no additional problem involving the definition of the wave function of the
ground state.

The substantiation of a possibility of the origination of the effective
electron-electron attraction in metals is given in study ~\cite{bib-7}.

There is, presumably, so far no phenomenological theory available on the
systems of such a kind.

\begin{acknowledgments}
We are grateful to Prof. A.A. Rukhadze for being helpful in the statement
of the problem.
\end{acknowledgments}

\appendix*
\section{}
Taking into account Eqs. (\ref{12}), it is possible to expect the fulfillment of
the following conditions:
\begin{eqnarray}
  && \chi _{0,{\bf q}}^{{\bf p},{\bf k}}  = \chi _{0,{\bf k} - {\bf p} - {\bf q}}^{{\bf p},{\bf k}} ,\chi _{0,{\bf q}}^{{\bf p},{\bf k}}  = \chi _{0, - {\bf q}}^{{\bf k},{\bf p}} ,  \cr
  && \chi _{1,{\bf q}}^{{\bf p},{\bf k}}  =  - \chi _{1,{\bf k} - {\bf p} - {\bf q}}^{{\bf p},{\bf k}} ,\chi _{1,{\bf q}}^{{\bf p},{\bf k}}  = \chi _{1, - {\bf q}}^{{\bf k},{\bf p}} ;  \cr
  && \sum\limits_{\bf z} {\chi _{0,{\bf z}}^{{\bf p},{\bf k} + {\bf q}} \chi _{0,{\bf q} - {\bf z}}^{{\bf k},{\bf p} + {\bf q}} }  = \frac{1}
{2}\left( {\delta ^0_{\bf q}  + \delta ^{\bf k}_{\bf p} } \right)
\cr
  && \sum\limits_{\bf z} {\chi _{1,{\bf z}}^{{\bf p},{\bf k} + {\bf q}} \chi _{1,{\bf q} - {\bf z}}^{{\bf k},{\bf p} + {\bf q}} }  = \frac{1}
{2}\left( {\delta ^0_{\bf q}  - \delta ^{\bf k}_{\bf p} } \right).
\lb{31}
\end{eqnarray}
The fulfillment of these conditions is conducive to that the
fermions commutation relationships for operators $C^ +$ and $C$ ,
as well, as their reciprocal transformation (\ref{13}), are
violated only beginning with 4-linear terms on the operators of
creation and annihilation. It is just this circumstance that is
employed in constructing the model Hamiltonian of the system.

\end{document}